\documentstyle[prb,aps,floats]{revtex} 
\begin{document} 
\input epsf
%\draft
\title{Kinetic equation approach to diffusive superconducting
hybrid devices} 
\author{T. H. Stoof and Yu. V. Nazarov} 
\address{Faculty of Applied Physics and Delft Institute for
Microelectronics and Submicron Technology (DIMES)\\
Delft University of Technology, Lorentzweg 1, 2628 CJ Delft,
The Netherlands}
%\date{\today}
\address{\rm  (Submitted to Physical Review B)}
\address{\mbox{ }}
%\maketitle 
\address{\parbox{14cm}{\rm \mbox{ }\mbox{ }\mbox{ }
%\begin{abstract}
We present calculations of the temperature-dependent electrostatic and
chemical potential distributions in disordered normal
metal-superconductor structures. We show that they differ appreciably
in the presence of a superconducting terminal and propose an experiment
to measure these two different potential distributions.
We also compute the resistance change in these structures due to a
recently proposed mechanism which causes a finite effect at zero temperature.
The relative resistance change due to this effect is of the order of the
interaction parameter in the normal metal.
Finally a detailed calculation of the resistance change due to the
temperature dependence of Andreev reflection in diffusive systems is
presented.
We find that the maximal magnitude due to this thermal effect is in general
much larger than the magnitude of the novel effect.
%\end{abstract}  
}}
\address{\mbox{ }\mbox{ }}
%\pacs{PACS numbers: 74.50.+r;74.80.Fp}
\twocolumn
\address{\parbox{14cm}{\rm PACS numbers: 74.50.+r;74.80.Fp}}
\maketitle

\section{Introduction} 
\label{introduction} 
Mesoscopic structures in which normal metal wires or semiconductors are 
attached to superconductors have received a fair amount of attention 
in the past few years. In particular devices known as Andreev interferometers, 
in which two superconducting terminals with different macroscopic phases 
are present, have been in the focus of interest. 
The conductance of these structures, in which electrons and 
holes undergo multiple Andreev reflection depends 
on the phase difference of the connected superconductors, hence the 
name Andreev interferometry. 
 
Since the prediction of Andreev reflection,\cite{andreev} the theory 
of charge transport through N-S junctions has been well
established.\cite{btk,lambert,beenakker} However, the practical 
implications of this phenomenon
for the sophisticated nano-structures that can nowadays be realized are 
not always clear. The reason for this is the coherent nature of multiple
Andreev reflection which determines the physical behavior of these
devices. 
These new technological developments resulted in the current revival of
the topic in mesoscopic physics. 
 
In the last few years a large number of Andreev interferometers 
have been studied both 
theoretically\cite{nanako,hekking,zaitsev,kadi,allsopp,circuit} and 
experimentally.\cite{vegvar,pothier,dimoulas,petrashov} 
Particularly the experiment of Ref.~\onlinecite{petrashov} motivated 
the research presented here. In this experiment, the resistance of a 
cross shaped diffusive normal metal was measured. The two branches of 
the cross perpendicular to the current path were in contact with a 
large superconducting loop. The phase difference between the
superconducting end points of the loop could be controlled by a 
small current through the loop or, alternatively, by applying 
a magnetic field. The resistance of the structure oscillated
non-harmonically as a function
of the phase difference by about 10\% of the normal state resistance. 
These results were unexpected because in the conventional theory of
the proximity effect, in which electron-electron interactions in the
normal metal region are disregarded, the zero-voltage, 
zero-temperature resistance of a diffusive metal is predicted to be 
phase-independent.\cite{artemenko,ns} Furthermore, the large amplitude
and the observed $2 \pi$ periodicity ruled out the possibility of
a weak localization effect, since the latter was predicted to show
a phase-dependence with a fundamental period of
$\pi$.\cite{spivak,altshuler} To this day, resistance oscillations with
$\pi$ periodicity remain unobserved.
 
Recently we proposed a new mechanism which provides a 
phase-dependent resistivity in a diffusive conductor at zero 
temperature.\cite{ns} 
This scheme takes into account the fact that the electron-electron 
interaction induces a weak pair potential in the normal metal. As a 
result, Andreev reflection occurs in the entire structure, 
rather than only at the N-S interfaces. This results in a phase-dependent 
resistance change which is proportional to the interaction parameter 
$\lambda$ in the normal metal and can be of either sign, depending on
the sign of $\lambda$. Although the experiment of
Ref.~\onlinecite{petrashov} 
could be explained in terms of the proximity effect 
theory and the results were shown to be caused by the finite temperature at 
which the experiments were performed,\cite{ns} it would be challenging
to observe the resistance oscillations predicted in Ref.~\onlinecite{ns}.
This would also be of practical interest since it would provide the means
to directly measure the interaction parameter in the normal metal.
However, besides the fact that electron-hole coherence influences
the resistance, it also manifests itself in a non-trivial distribution of
the electrostatic and chemical potentials in the structure as we will show
below.
 
The remainder of this paper is organized as follows.
In section~\ref{potsec} we briefly discuss the influence of phase 
coherence on transport properties and potential distributions in small
diffusive structures.
Section~\ref{theory} contains the theoretical foundation of our
calculations. We first review the relevant techniques of the Keldysh
formalism for diffusive superconductors and then derive the equations
for the Green functions and distribution functions which determine the
electric transport properties of the system. The next three sections
are devoted to several applications of the theory.
We first calculate the temperature-dependent electrostatic and chemical
potential distributions in a simple one dimensional structure 
in section~\ref{distribution}.
A second application is presented in section~\ref{results}, where we
calculate the resistance change at zero temperature due to the induced
pair potential in the normal metal region for two experimentally relevant
geometries.
The third and last application is discussed in section~\ref{thermal}.
There we focus our attention on the influence of a finite temperature on
the resistance in these structures. Some of the results in sections
\ref{results} and \ref{thermal} were published in a preliminary form in
Ref.~\onlinecite{ns}. However, here we additionally give a detailed
description of the performed calculations.
We summarize our conclusions in section~\ref{end}.
 
\section{Coherence effects in ultrasmall disordered structures} 
\label{potsec}

Owing to the advance in nanofabrication techniques over the past years,
the fabrication of hybrid metallic superconducting structures with a
characteristic size of a few microns or less has nowadays become
possible. If these small structures are at a sufficiently low
temperature, the quasiparticles in the metal can no longer penetrate into
the superconductor due to the large superconducting gap. As a result,
the lowest order process that determines the resistance of the
system is Andreev reflection, in which an electron is reflected as a hole
or, alternatively, in which an electron pair enters the
superconductor. This reflection causes electrons and holes in the
diffusive
metal to be phase coherent over distances of the order of
$\xi=\sqrt{{\cal D}/T} \gg l$, where ${\cal D}=\frac{1}{3} v_{\rm F} l$
is the diffusion constant and $l$ is the elastic mean free path.

This phase coherence between electrons and holes drastically alters the
physics of transport through such systems. The most striking feature
is that the electrostatic potential and nonequilibrium chemical
potential are no longer distributed linearly through the sample.
The nonlinearity of the electrostatic potential implies a nonuniform
resistivity distribution and consequently a non-local resistance of the 
structure. This nonlocality is a fundamental feature
of the coherent nature of Andreev reflection.
Moreover, at finite temperatures the transport properties of the system
cease to be distributed uniformly over all energies. 
Hence, a calculation of these quantities,
to which the main part of this article will be devoted, must first
consider them at each energy individually and then integrate over all
energies.

Another manifestation of the phase coherence in the normal
metal is the difference in the distribution functions of the electrostatic
and chemical potentials. In a normal system both would be equal, but
this changes when one of the leads is brought into the superconducting
state. Whereas the former is simply determined by the
distribution of charge in the system, the latter can only be defined for
small perturbations from equilibrium, i.e. when the quasiparticle energies
are much smaller than the superconducting gap. To show how this definition
comes about we consider the normal current through a disordered N-I-S
junction. Zhou, Spivak, and Zyuzin showed that for small quasiparticle
energies $\varepsilon$ this current can be written in the following
way:\cite{zhou}
\begin{equation}
\label{noneqcur}
j(\varepsilon,x_{\rm N})= t \left\{
f_{\rm T}(\varepsilon,x_{\rm S}) - f_{\rm T}(\varepsilon,x_{\rm N})
\right\} F(\varepsilon,x_{\rm N},x_{\rm S}),
\end{equation}
where $t$ is the transparency of the tunnel barrier, $x_{\rm N}$ and
$x_{\rm S}$ denote the normal-metal side and superconducting side of the
barrier, $F$ is some function of $\varepsilon$, $x_{\rm N}$ and
$x_{\rm S}$, and $f_{\rm T}$ is the nonequilibrium
distribution function. This equation shows that at low temperatures
the nonequilibrium chemical potential can be associated with the
distribution function and consequently that it is a measurable quantity.
As will be discussed in section~\ref{distribution}, the electrostatic
potential decreases faster
than linear in the vicinity of the superconductor due to the decreased
density of states near the N-S interface. In contrast, the chemical
potential changes only a little in the presence of a superconducting
terminal and consequently the ratio of the electrostatic and chemical
potential vanishes near the superconductor.

\begin{figure}
\epsfxsize=8cm \epsfbox[18 300 577 750]{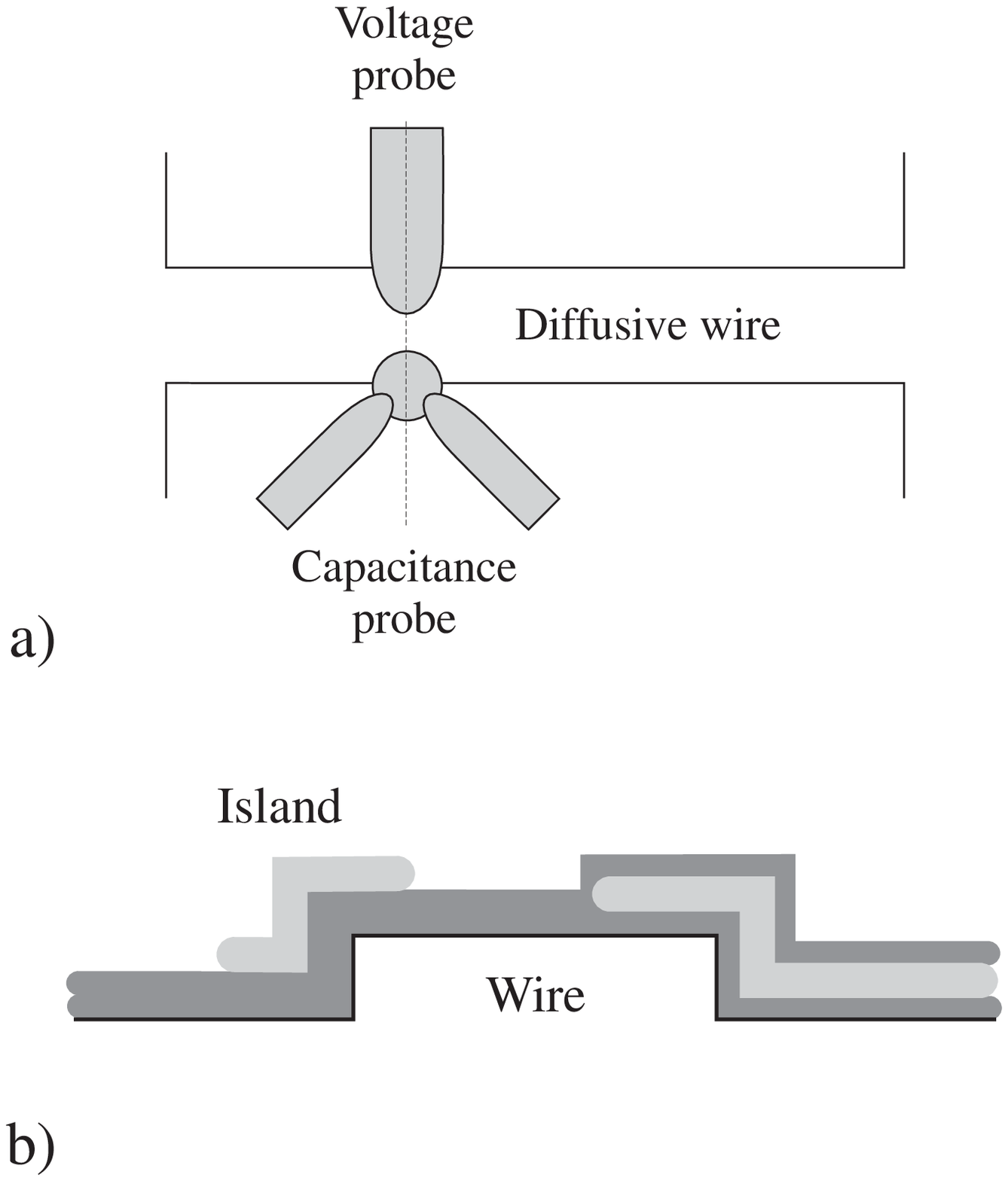}
\caption{Schematic setup of an experiment to measure the electrostatic 
and chemical potential: (a) Top view. (b) Cross section along the 
dashed line in a). For details see text.} 
\label{fig:exp} 
\end{figure}

A possible experimental setup to measure the  
difference between the electrostatic and chemical potential is drawn 
in Fig.~\ref{fig:exp}.
The figure shows a diffusive wire connecting  
two reservoirs (not shown) in the presence of two different probes.
The voltage probe 
measures the chemical and the capacitance probe the electrostatic
potential. 
As shown in Fig.~\ref{fig:exp}(b), the voltage probe consists simply of a 
metallic lead separated from the wire by a thin oxide layer. The latter 
is indicated by the dark shaded region. The capacitance probe is slightly 
more complicated. In principle one could use a single metallic gate 
separated by a thick oxide layer to reduce tunneling from lead to wire. 
However, because such a gate would not only couple capacatively to the 
charge in the wire directly underneath the tip but also strongly to 
the surroundings, we propose a slightly different method: A small 
metallic island is deposited on the edge of the wire as indicated in 
the figure. This island is weakly coupled to two extra leads through 
which a current can flow. In this case the electrostatic potential 
capacatively induces a charge on the island, thus very sensitively 
changing the measured Coulomb threshold. Both probes can be 
calibrated since they should measure the same potentials if both 
reservoirs are in the normal state. If one of the reservoirs is brought 
into the superconducting state, both potentials should change.
If one would attach different probes along the wire it would 
be possible to measure the potential landscapes in the wire.
In section~\ref{distribution} we will return to this subject in a more
quantitative fashion, but we will now first discuss the necessary theory.
 
\section{Method} 
\label{theory} 
 
\subsection{Kinetic equations for the distribution functions in the dirty
limit} 
\label{kinetic} 
To describe the system we use the nonequilibrium Green function 
method first derived by Keldysh\cite{keldysh} and later 
further developed for superconductivity by Larkin and 
Ovchinnikov.\cite{larkin} Although this framework is rather formal
it has one big advantage over e.g. the scattering approach. As was
mentioned above the resistance of the structure is nonlocal. However,
using the Keldysh technique it is possible to express scattering
processes in the structure as well as other relevant physical quantities,
like the resistance, in terms of local Green functions in coincident
points.
This property of the formalism simplifies the calculations considerably.
To establish notation and to remind the reader of the basic theory we
briefly review the essential ingredients of the Keldysh formalism and the
quasiclassical approximation for diffusive superconductors. For more
extensive
reviews we refer to e.g. Refs.~\onlinecite{rev1} and~\onlinecite{rev2}. 

In this formalism the Green function is given by the  
$(4 \times 4)$ matrix 
\begin{equation} 
\label{green} 
\check{G}= 
\left( 
\begin{array}{cc} 
\hat{G}^{\rm A} & \hat{G}^{\rm K}\\ 
0 & \hat{G}^{\rm R} 
\end{array} 
\right), 
\end{equation} 
where $\hat{G}^{\rm A},\hat{G}^{\rm R}$, and $\hat{G}^{\rm K}$ are the 
advanced, retarded, and Keldysh Green function which are 
$(2 \times 2)$ matrices in Nambu space given by\cite{agd}
\begin{equation} 
\label{ark} 
\hat{G}^{\rm A}(1,1')= 
\left( 
\begin{array}{cc} 
G^{\rm A}(1,1') & F^{\rm A}(1,1') \\ 
F^{A \dagger}(1,1') & -G^{\rm A}(1',1) 
\end{array} 
\right), 
\end{equation} 
and analogous equations for $\hat{G}^{\rm R}$ and $\hat{G}^{\rm K}$. 
Throughout this paper the symbol "check" will be used to denote
$(4 \times 4)$ matrices and "hat" for $(2 \times 2)$ matrices.
The normal  and anomalous Green function are given by: 
\begin{mathletters} 
\begin{eqnarray} 
\label{defs} 
G^{\rm R}(1,1')&=&-i \theta(t_{1}-t_{1'}) 
\langle[\psi(1),\psi^{\dagger}(1')]_{+}\rangle, \\ 
G^{\rm A}(1,1')&=&i \theta(t_{1'}-t_{1}) 
\langle[\psi(1),\psi^{\dagger}(1')]_{+}\rangle, \\ 
G^{\rm K}(1,1')&=&-i \langle[\psi(1),\psi^{\dagger}(1')]\rangle, \\ 
F^{\rm R}(1,1')&=&-i \theta(t_{1}-t_{1'}) 
\langle[\psi(1),\psi(1')]_{+}\rangle, \\ 
F^{\rm A}(1,1')&=&i \theta(t_{1'}-t_{1}) 
\langle[\psi(1),\psi(1')]_{+}\rangle, \\ 
F^{\rm K}(1,1')&=&i \langle[\psi^{\dagger}(1),\psi^{\dagger}(1')]\rangle, 
\end{eqnarray} 
\end{mathletters} 
where $\psi(1)=\psi(t_{1},{\bf r}_{1})$ is the electron field operator. 
 
We proceed by introducing the center-of-mass and relative 
coordinates ${\bf r}=\frac{1}{2}({\bf r}_{1}+{\bf r}_{1'})$ and 
${\bf r}'={\bf r}_{1}-{\bf r}_{1'}$ and by Fourier transforming the 
Green function with respect to the relative coordinate: 
$\check{G}({\bf r},{\bf p})= \int d{\bf r}' \exp({-i{\bf p}{\bf r}'})\; 
\check{G}({\bf r}+\frac{1}{2}{\bf r}',{\bf r}-\frac{1}{2}{\bf r}')$. 
We apply the quasiclassical approximation, which is based on 
the fact that the Fermi energy in the system is much larger than all 
other energy scales. This means that all relevant physical quantities
vary spatially on a length scale that is much larger 
than the Fermi wavelength. In this case it is useful to introduce 
the so-called quasiclassical Green function $\check{g}$ which 
is integrated over $\xi_{\bf p}=\frac{{\bf p}^{2}}{2m}-\mu$: 
\begin{equation} 
\label{quasig} 
\check{g}({\bf r},\hat{p},t_{1},t_{1'})= 
\frac{i}{\pi} \int d\xi_{\bf p} \check{G}({\bf r},{\bf p},t_{1},t_{1'}). 
\end{equation} 
Here $\hat{p}$ in the left hand side denotes the fact that the momentum 
dependence of the quasiclassical Green function is restricted to 
dependence on the direction of ${\bf p}$ only. In this approximation 
the magnitude of the momentum is fixed at $|{\bf p}|= 
p_{\rm F}$. This quasiclasissical Green function satisfies the
normalization condition
\begin{equation}
\label{normalization}
\int dt_{1''} \check{g}(t_{1},t_{1''}) \check{g}(t_{1''},t_{1'})=
\check{1} \delta(t_{1}-t_{1'}). 
\end{equation}

In the case of a superconductor with short elastic mean free path, i.e.
in the diffusive regime, it is feasible to expand the Green function to
first order in spherical harmonics:\cite{usadel} 
\begin{equation} 
\label{spherical} 
\check{g}=\check{g}_{s}+{\bf p} {\bf \check{g}}_{p},~~~ 
{\bf p} {\bf \check{g}}_{p} \ll \check{g}_{s}, 
\end{equation} 
where the functions $\check{g}_{s}$ and ${\bf \check{g}}_{p}$ no 
longer depend on the direction of ${\bf p}$. Using the normalization  
condition (\ref{normalization}) we find an expression for 
${\bf \check{g}}_{p}$, which is then substituted back into
Eq.~(\ref{spherical}). The thus obtained Green function is then
averaged over all angles of ${\bf p}$.
In the stationary case, the Green function depends on the 
time difference $\tau = t_{1}-t_{1'}$ only. Performing the Fourier 
transform with respect to this time difference, we obtain the equation of 
motion for the Green function $\check{g}_{\varepsilon} = \int d \tau 
\check{g}_{s}(\tau) \exp(i \varepsilon \tau)$ (we drop the index "s" from 
now on)\cite{volkov} 
\begin{mathletters}
\begin{eqnarray}
\label{dirtyeq2} 
-{\cal D} {\bf \nabla} \left( \check{g}_{\varepsilon} 
{\bf \nabla} \check{g}_{\varepsilon} \right) + 
i[\check{H},\check{g}_{\varepsilon}] +
i[\check{\Sigma},\check{g}_{\varepsilon}]=0, \\
\check{g}^{2}_{\varepsilon}=\check{1}, 
\end{eqnarray} 
\end{mathletters}
where ${\cal D}=\frac{1}{3} v_{\rm F} l$ is the diffusivity,
$l$ is the elastic mean free path,
$\check{H}=e \varphi \check{1} + \varepsilon \check{\sigma}_{z} 
-\check{\Delta}$ and $\check{\Delta}$ and $\check{\sigma}_{z}$ are
given by:
\begin{equation} 
\label{delta} 
\check{\Delta}= 
\left( 
\begin{array}{cc} 
\hat{\Delta} & 0\\ 
0 & \hat{\Delta} 
\end{array} 
\right) 
,  
\hat{\Delta}= 
\left( 
\begin{array}{cc} 
0 & \Delta\\ 
-\Delta^{*} & 0 
\end{array} 
\right), 
\end{equation} 
where $\Delta$ is the pair potential in the metal and 
\begin{equation} 
\check{\sigma}_{z}= 
\left( 
\begin{array}{cc} 
\hat{\sigma}_{z} & 0\\ 
0 & \hat{\sigma}_{z} 
\end{array} 
\right) 
,  
\hat{\sigma}_{z}= 
\left( 
\begin{array}{cc} 
1 & 0\\ 
0 & -1 
\end{array} 
\right). 
\end{equation} 
In Eq.~(\ref{dirtyeq2}) elastic non spin-flip impurity scattering has been
taken into account in the Born approximation, causing the presence of the
elastic mean free path $l$ in the diffusion constant.\cite{rev2}
Hence, the self energy matrix $\check{\Sigma}$ in Eq.~(\ref{dirtyeq2}),
which has the same structure as the Green function (\ref{green}), takes
into account processes like spin-flip scattering and the inelastic
scattering of electrons with phonons and (magnetic) impurities. 

The general expression for the electrostatic potential $\varphi$ in
Eq.~(\ref{dirtyeq2}), which follows from electroneutrality in the metal,
is\cite{rev2,zhou}
\begin{equation}
\varphi(x)=-\frac{1}{8e} \int_{-\infty}^{\infty} d\varepsilon
\;\mbox{Tr}\; \hat{g}^{\rm K}_{\varepsilon}(x).
\end{equation}
Throughout this paper the electrostatic potential is assumed to be
time-independent. 

The advanced and retarded Green functions determine the dispersion
of the quasiparticles. However, to solve a transport problem
we need to know how the energy spectrum is filled by extra quasiparticles
when the system is driven out of equilibrium.
This is determined by the Keldysh component $\hat{g}^{\rm K}$ of
$\check{g}$, which can be expressed in the advanced and retarded ones
using two distribution functions:\cite{larkin,rev1,rev2} 
\begin{mathletters} 
\label{ansatz} 
\begin{eqnarray} 
\hat{g}_{\varepsilon}^{\rm K}&=&\hat{g}_{\varepsilon}^{\rm R} \hat{f} - 
\hat{f} \hat{g}_{\varepsilon}^{\rm A}\\ 
\hat{f}&=&f_{\rm L} \hat{1} + f_{\rm T} \hat{\sigma}_{z}. 
\end{eqnarray} 
\end{mathletters} 
In a spatially slowly varying electro-magnetic field, the equations for 
the two distribution functions are (dropping collision integrals because
they account for inelastic scattering processes, and time derivatives
because we seek to find stationary solutions only) 
\begin{mathletters} 
\label{distfunc} 
\begin{eqnarray} 
{\cal D} {\bf \nabla} \mbox{Tr} \left\{ 
{\bf \nabla} f_{\rm L} (\hat{1}-\hat{g}_{\varepsilon}^{\rm R} 
\hat{g}_{\varepsilon}^{\rm A}) \right\}+
{\cal D} {\bf \nabla} \left( f_{\rm T} j_{\varepsilon} \right)&=&0\\
\nonumber 
{\cal D} {\bf \nabla} \mbox{Tr} \left\{ {\bf \nabla} f_{\rm T} (\hat{1} -
\hat{\sigma}_{z} \hat{g}_{\varepsilon}^{\rm R} \hat{\sigma}_{z}
\hat{g}_{\varepsilon}^{\rm A})\right\}+
{\cal D}{\bf \nabla}f_{\rm L}j_{\varepsilon}&+&\\
2i f_{\rm T} \mbox{Tr}\left\{ (\hat{g}_{\varepsilon}^{\rm R}+ 
\hat{g}_{\varepsilon}^{\rm A}) \hat{\Delta} \right\}&=&0 
\end{eqnarray} 
\end{mathletters} 
where $j_{\varepsilon}=\mbox{Tr} \hat{\sigma}_{z} \{  
\hat{g}_{\varepsilon}^{\rm R} \hat{\partial}
\hat{g}_{\varepsilon}^{\rm R}- 
\hat{g}_{\varepsilon}^{\rm A} \hat{\partial}
\hat{g}_{\varepsilon}^{\rm A} \}$.

To close the set of equations we finally need an equation for the pair 
potential in the normal metal region. This expression for $\Delta$ can be
derived from the selfconsistency relation:
\begin{eqnarray}
\label{gapmatrix}
\nonumber
\hat{\Delta}({\bf r}) &=& \frac{\lambda}{4 i} \int d\varepsilon \; 
\left\{ \hat{g}^{\rm K}_{\varepsilon}({\bf r},\hat{p},\tau)
\right\}_{\mbox{o.d}},\\&=&
\frac{\lambda}{4 i} \int d\varepsilon \;
\tanh\left(\frac{\varepsilon}{2 T}\right) \left\{
\hat{g}^{\rm R}_{\varepsilon}-
\hat{g}^{\rm A}_{\varepsilon} \right\}_{\mbox{o.d}},
\end{eqnarray}
where $\lambda=g N(0)$ is the interaction parameter, $g$, times the
density of states at the Fermi level, $N(0)$. The subscript o.d.
denotes the off-diagonal part.

This concludes the derivation of the distribution functions. We now
have a closed system of equations that in principle must be solved 
selfconsistently. In the next section we will discuss specific
circumstances that allow for a simplification of the equations,
enabling us to solve them perturbatively.

\subsection{Approximations} 
\label{interaction} 
In this section we discuss the assumptions and approximations that
enable us to simplify the theory. We subsequently derive the final
set of equations that we will use.

We start by noting that Eq.~(\ref{dirtyeq2}) for the Green function
still contains the self energy matrix $\check{\Sigma}$.
However, because we are only interested in the case where the phase 
breaking length is much larger than the system size, it is reasonable 
to disregard inelastic scattering processes and hence, we neglect  
$\check{\Sigma}$ from now on. Eq.~(\ref{dirtyeq2}) then reduces 
to\cite{ns} 
\begin{equation} 
\label{simpel} 
{\cal D} {\bf \nabla} \left( \check{g}_{\varepsilon} 
{\bf \nabla} \check{g}_{\varepsilon} \right) - 
i[\check{H},\check{g}_{\varepsilon}]=0. 
\end{equation}
We now have equations for the diagonal components
$\hat{g}_{\varepsilon}^{\rm A}$ and $\hat{g}_{\varepsilon}^{\rm R}$ of the
Green function~(\ref{green}). 

We parametrize the advanced Green function in the following way:
\begin{equation}
\label{param}
\hat{g}^{\rm A}_{\varepsilon} = \left(
\begin{array}{cc}
\cos \theta & i e^{i \phi} \sin \theta \\
-i e^{-i \phi} \sin \theta & - \cos \theta
\end{array} \right),
\end{equation}
thus ensuring that $\hat{g}_{\varepsilon}^{2}=\hat{1}$.
In general $\phi$ and $\theta$ are complex and depend on energy and
position. In a structure with two superconducting terminals,
$\hat{g}^{\rm A}_{\varepsilon}$ will depend on the phase {\em difference}
$\phi_{1}-\phi_{2}$ between the two superconductors. However, if only
one superconducting reservoir is present, the resistance of the structure
will not depend on the absolute phase and we can put $\phi=0$.

In the case of sufficiently small quasiparticle, thermal and Thouless
energies; $\varepsilon, k_{\rm B}T,{\cal D}/L^{2} \ll \Delta_{\rm S}$,
where $\Delta_{\rm S}$ is the energy gap in the superconductor, the
advanced Green function can be written in the following way:\cite{larkin}
\begin{equation}
\label{gapprox}
\hat{g}^{\rm A}_{\varepsilon} =
\frac{-1}{\sqrt{(\varepsilon-i\delta)^{2}-|\Delta|^{2}}}
\left(
\begin{array}{cc}
\varepsilon & \Delta \\
-\Delta^{*}& - \varepsilon
\end{array}
\right),
\end{equation}
where $\delta$ is an infinitesimally small positive number. Using this
representation for $\hat{g}^{\rm A}_{\varepsilon}$ it is easy to
derive boundary conditions for Eq.~(\ref{simpel}). In a normal
reservoir $\Delta=0$ and $\hat{g}_{\varepsilon}^{A}=
-\hat{\sigma}_{z}$. In a superconducting terminal having phase $\phi$,
$\Delta=|\Delta_{\rm S}| e^{i \phi}$ and the Green function
satisfies $\hat{g}^{\rm A}_{\varepsilon}=\hat{\sigma}_{x} \sin \phi +
\hat{\sigma}_{y} \cos \phi$.

It is also possible to simplify Eq.~(\ref{distfunc}) for the distribution 
functions considerably: In the case of a negligible supercurrent
$I_{s} = \int j_{\varepsilon} d\varepsilon$ the equations for the two
distribution functions decouple, reducing $f_{\rm L}$ to
its equilibrium value $f_{\rm L}=\tanh(\frac{\varepsilon}{2 T})$ and
leaving us with a single equation for $f_{\rm T}$ which can be cast into
the form of a diffusion equation:\cite{ns}
\begin{equation}
\label{diffeq}
{\bf \nabla} \left( D(\varepsilon,{\bf r}) {\bf \nabla}
f_{\rm T}(\varepsilon,{\bf r}) \right) - \gamma(\varepsilon,{\bf r})
f_{\rm T}(\varepsilon,{\bf r})=0,
\end{equation}
where the first term describes diffusion of quasiparticles with
an effective diffusion coefficient
\begin{eqnarray}
\label{diffcoef}
\nonumber
D(\varepsilon,{\bf r})&=&\frac{{\cal D}}{4}
\mbox{Tr}\{\hat{1}-\hat{\sigma}_{z} \hat{g}_{\varepsilon}^{\rm R}
\hat{\sigma}_{z} \hat{g}_{\varepsilon}^{\rm A} \},\\
&=&\frac{{\cal D}}{8} \mbox{Tr}
\{{(\hat{g}^{\rm A}_{\varepsilon}+
\hat{g}^{A \dagger}_{-\varepsilon})^{2}}\},
\end{eqnarray}
that is modified by the penetrating superconductivity. Here we have
used the identity $\hat{g}^{\rm R}_{\varepsilon}=-\hat{\sigma}_{z}
\hat{g}^{\rm A}_{-\varepsilon}\hat{\sigma}_{z}$, which relates the
advanced
and retarded Green functions.
The second term describes absorption of quasiparticles in the
superconducting condensate with a coefficient
$\gamma = \frac{i}{2} \mbox{Tr}
\{ (\hat{g}_{\varepsilon}^{\rm R}+ \hat{g}_{\varepsilon}^{\rm A})
\hat{\Delta} \}$.
In the absence of external fields $\gamma$ is proportional to the
local value of the pair potential, since in that case we can always
choose $\Delta$ to be real, and we obtain
\begin{equation}
\label{gamma1}
\gamma({\bf r},\varepsilon)=-\frac{i}{2}
\Delta({\bf r}) \mbox{Tr}\{ i\hat{\sigma}_{y}
(\hat{g}_{-\varepsilon}^{\rm A}+
\hat{g}_{\varepsilon}^{\rm A})\}.
\end{equation}
The boundary conditions for Eq.~(\ref{diffeq}) follow from expanding
$\tanh(\frac{\varepsilon+eV}{2T})$ to first order in $V$. This determines
the boundary condition for $f_{\rm T}$. In a normal reservoir that is
biased at a (small) voltage $V$ with respect to a superconducting lead,
the distribution function is
$f_{\rm T}=(\frac{eV}{2T})\cosh^{-2}(\frac{\varepsilon}{2T})$.

In most theoretical approaches, see e.g.
Refs.~[\onlinecite{lambert,beenakker,circuit,volkov}],
electron-electron interactions in the normal metal are disregarded,
leading to $\Delta, \gamma = 0$.
However, as shown in Ref.~\onlinecite{ns}, including the effect of
these interactions produces a change in the resistance. We thus need
Eq.~(\ref{gapmatrix}) for $\Delta$, which after a straightforward
calculation can be rewritten as:
\begin{equation}
\label{gap}
\Delta = \frac{\lambda}{8i} \int d\varepsilon \; 
\tanh\left(\frac{\varepsilon}{2 T}\right) \mbox{Tr} \left\{ 
i \hat{\sigma}_{y}
(\hat{g}^{\rm A}_{\varepsilon}-\hat{g}^{\rm R}_{\varepsilon}) \right\}.
\end{equation}
We now have all the necessary ingredients to calculate the various 
non-equilibrium transport properties of the system. The next section 
will be devoted to two of these properties, namely the electrostatic and
chemical potential distributions.

\section{Electrostatic versus chemical potential} 
\label{distribution}

\begin{figure}
\epsfxsize=7.8cm \epsfbox[-75 75 577 800]{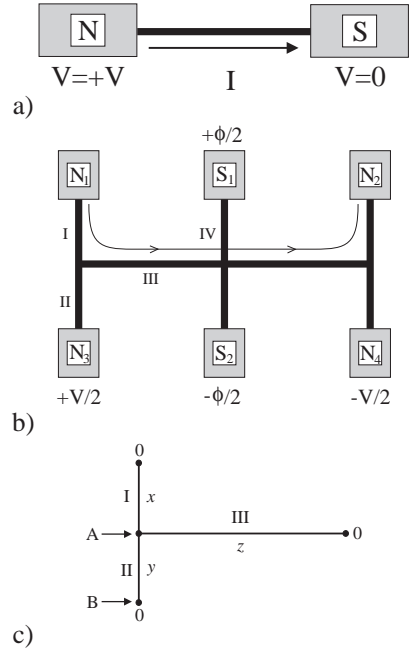}
\caption{(a) The simplest possible system, consisting of a diffusive
normal metal wire of length $L$, that is connected on the left to
a normal reservoir and to the right to a superconducting one. (b) An
example of a more complicated structure. The branches I, II and IV have
length $L'$, branch III has length $L$. For further details see text.
(c) The coordinates on the branches. The origins are indicated for each
branch.}
\label{fig:system}
\end{figure}

As a first example of the theory of section~\ref{theory} 
we calculate the electrostatic and the chemical potential
in the 1D wire of Fig.~\ref{fig:system}(a). As was shown above, the former
is determined by the distribution of electric charge in the wire.
The latter determines the magnitude of the current that flows through
the sample.
The system consists of a diffusive normal metal
wire of length $L$ attached on the left to a normal metal reservoir
and on the right to a superconducting terminal. The normal lead is
biased at a small voltage $V$ with respect to the superconductor.
The normal reservoir is situated at $x=0$.

\begin{figure}
\epsfxsize=8cm \epsfbox[18 300 577 750]{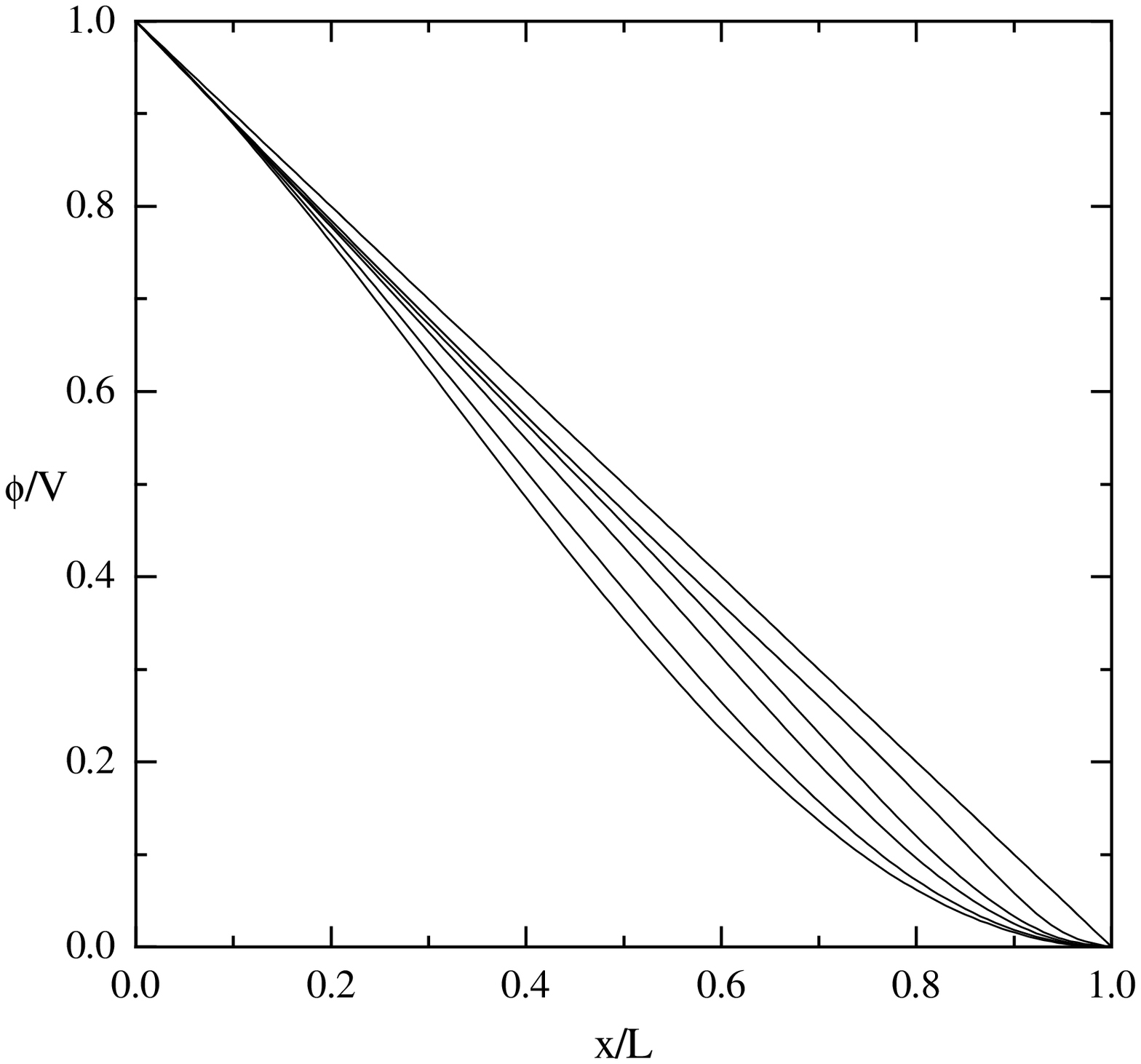}
\caption{Electrical potential distribution in the wire of
Fig.~\protect\ref{fig:system}(a) as
a function of temperature. Going from bottom to top, the curves
correspond to values of $L/\xi=0$, 1.0, 1.5, 2.0, 4.0, and $\infty$.
The temperature is proportional to $(L/\xi)^{2}$.}
\label{fig:pot}
\end{figure}

Using Eq.~(\ref{ansatz}) for the Keldysh component and Eq.~(\ref{param})
for the Green function we obtain
\begin{equation}
\varphi(x)=\frac{1}{e} \int_{0}^{\infty} d\varepsilon\;
f_{\rm T}(x,\varepsilon) \cos(\theta(x,\varepsilon)).
\end{equation}
In Fig.~\ref{fig:pot} we have calculated the potential distribution
in the normal metal for different values of $L/\xi$, i.e. for 
different temperatures.

In the limiting cases of low and high temperatures, the potential
can be calculated analytically:
\begin{mathletters}
\begin{eqnarray}
\label{analpot}
\varphi(x)=V \left(1-\frac{x}{L}\right)
\cos\left({\frac{\pi x}{2 L}}\right),~~T \rightarrow 0,\\
\varphi(x)=V \left(1-\frac{x}{L}\right),~~\frac{{\cal D}}{L^{2}}
 \ll k_{\rm B}T \ll \Delta_{\rm S}.
\end{eqnarray}
\end{mathletters}
Fig.~\ref{fig:pot} shows that the potential distribution changes with
temperature from a non trivial one which is influenced strongly by the
penetrating superconductivity to the expected linear dependence for
high temperatures (but still $k_{\rm B} T \ll \Delta_{\rm S}$). This
behavior is caused by the fact that the density of states vanishes in the
vicinity of the superconductor. Hence, the charge distribution which
causes the electrostatic potential also vanishes in this region.
The most important consequence of the nonlinear voltage distribution
across the sample is the fact that the resistance at a certain point
is no longer local, but depends on the distribution of resistivity
in the entire structure. This is a direct consequence of the 
coherent nature of Andreev reflection as was discussed above.

\begin{figure}
\epsfxsize=8cm \epsfbox[18 330 577 750]{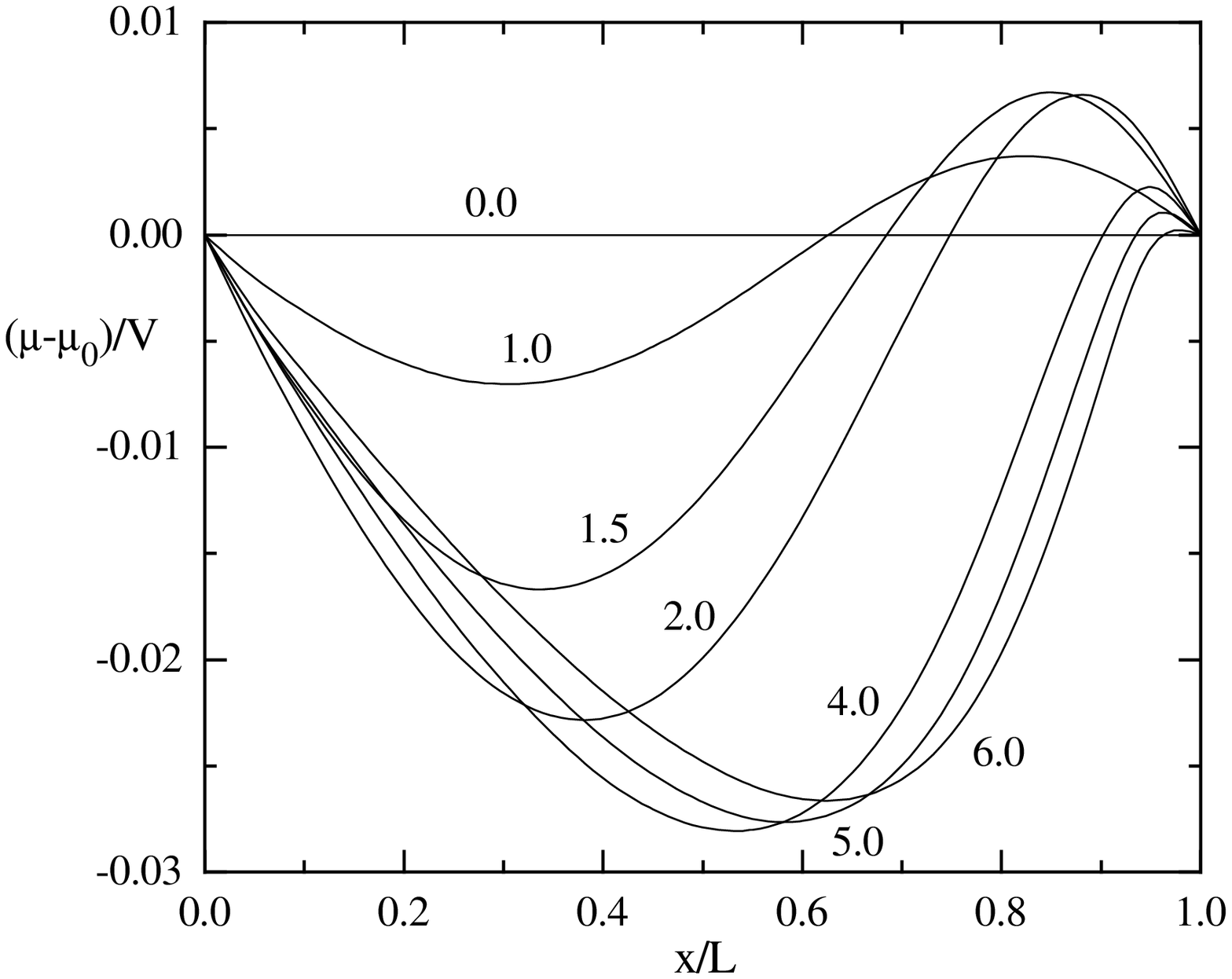}
\caption{Change in the chemical potential for the wire of
Fig.~\protect\ref{fig:system}(a) relative to the (linear) zero temperature
distribution as a function of temperature. The labels near the curves
indicate the value of $L/\xi$.}
\label{fig:mu}
\end{figure}

The chemical potential, which is simply proportional to the energy
integrated distribution function, is much less sensitive to changes in
temperature. The zero and high temperature distributions are the
same and are given by
\begin{equation}
\label{mu0}
\mu_{0}(x)=V \left( 1-\frac{x}{L} \right).
\end{equation}
In Fig.~\ref{fig:mu} we have plotted the deviation of the chemical
potential from this zero temperature solution.
First of all
we note that the change is very small. The maximum change at first
increases rapidly with increasing temperature. However, beyond a
certain temperature, $L/\xi \approx 4$, the maximum starts 
decreasing slowly. For high temperatures, the solution returns to 
the linear distribution. Thus we see that the electrostatic and chemical
potential are only equal in the case of high temperatures, when both
reduce to the trivial linear dependence that is also exhibited in the
case of two normal terminals.

\begin{figure}
\epsfxsize=8cm \epsfbox[18 330 577 750]{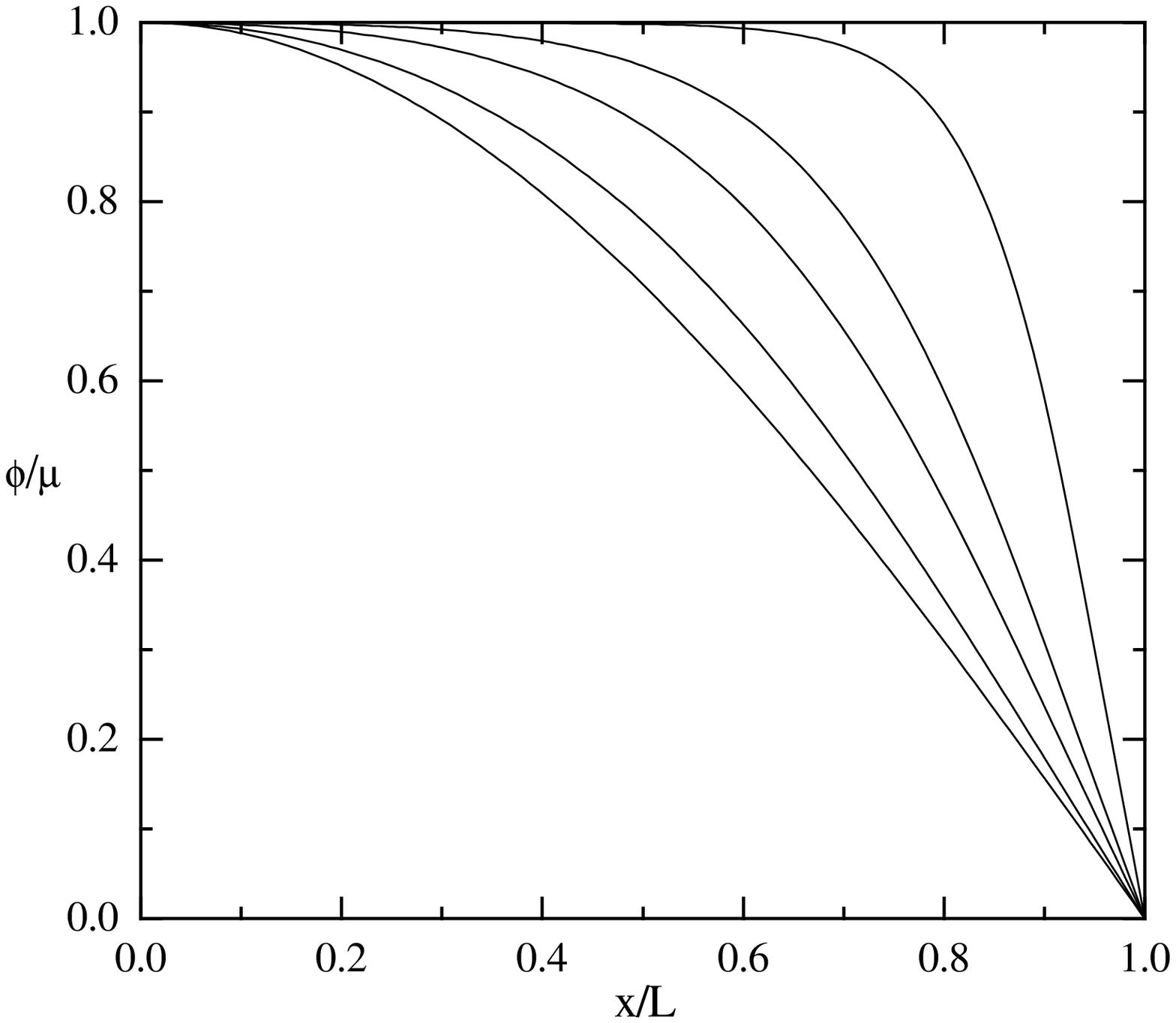}
\caption{The ratio of electrostatic and chemical potential as a function
of temperature. Going from bottom to top, the curves correspond to values
of $L/\xi=0$, 1.0, 1.5, 2.0, and 4.0.}
\label{fig:ratio}
\end{figure}

To indicate more clearly the difference between electrostatic and chemical
potential, we have plotted the ratio of the two for different
temperatures in Fig.~\ref{fig:ratio}.
The electrostatic and chemical
potential differ most at low temperatures and near the superconductor. 
However, in the vicinity of the normal reservoir they are almost equal
and the extent into the wire at which they are equal increases at higher
temperatures. In the high temperature limit they both reduce to the same
linear distribution, as was shown above.

\section{The interaction effect}
\label{results}

\subsection{The resistance of a 1D wire} 
\label{simple} 

In this section we calculate the resistance change at zero temperature
due to the penetration of the pair potential into the normal metal region.
We first consider the wire of Fig.~\ref{fig:system}(a) and then address
more general geometries.
In order to simplify the equations later on, the superconducting end
of the wire is now located at $x=0$ and the normal end at $x=L$.

We first solve Eq.~(\ref{simpel}) for $\hat{g}^{\rm A}$ numerically to
zeroth order in $\Delta$ and $\varphi$:
\begin{equation}
\label{argfeq}
{\cal D} {\bf \nabla} \left( \hat{g}^{\rm A}_{\varepsilon} 
{\bf \nabla} \hat{g}^{\rm A}_{\varepsilon} \right) - 
i\varepsilon[\hat{\sigma}_{z},\hat{g}^{\rm A}_{\varepsilon}]=0, 
\end{equation}
where $\varphi=0$ (zero voltage limit) and $\Delta$ has been disregarded
since it is small in the normal metal region because of its
proportionality to the interaction parameter $\lambda$.
We then solve Eq.~(\ref{diffeq}) to first order in $\gamma$. At zero
temperature, $D(\varepsilon,x)={\cal D}$, and we write $f_{\rm T}=f_{0}+
f_{1}$
with $f_{1} \ll f_{0}$. Then $f_{0} = c(\varepsilon) \frac{x}{L}$ and
$f_{1}''=\frac{\gamma}{{\cal D}} f_{0}$, with $c(\varepsilon)=
\frac{eV}{4T}\cosh^{-2}(\frac{\varepsilon}{2T})$. Using the fact that
$\int_{0}^{L} f_{1}'(x') dx'=0$ ($f_{1}(0)=f_{1}(L)=0$) we find
\begin{mathletters}
\begin{eqnarray}
\label{effen}
f_{0}'(L)&=&\frac{c(\varepsilon)}{L},\\f_{1}'(L)&=&
\frac{c(\varepsilon)}{{\cal D}L^{2}} \int_{0}^{L} x^{2} \gamma(x) dx.
\end{eqnarray}
\end{mathletters}
The current in a normal piece of metal is proportional to the local
gradient of $f$. Since everywhere in the diffusive wire there is an
induced pair potential it is not possible to calculate the current in
this way somewhere in the middle of the wire. We must calculate it at
the normal reservoir, where the pair potential is forced to vanish.
The current flowing out of the normal reservoir (and hence the conductance
$G$ of the system) is therefore proportional to $f'(L)$. 
Noting that $\frac{\delta R}{R}=-\frac{\delta G}{G}$ we obtain for the
relative resistance change:
\begin{equation}
\label{relres}
\frac{\delta R}{R} = -\frac{f_{1}'(L)}{f_{0}'(L)}=
-\frac{1}{{\cal D} L}\int_{0}^{L} dx\;x^{2}\;\gamma(x),
\end{equation}
where $\gamma$ is given by
\begin{equation}
\label{gamma}
\gamma(x) = 2 \Delta(x) \sin(\theta(\varepsilon=0,x)).
\end{equation}
Thus the relative resistance change is proportional to the interaction
parameter $\lambda$ and its sign depends on the sign of $\lambda$.
Here we use the convention that $\lambda$ is positive for attractive
effective interactions in the metal. Furthermore, the resistance change
depends sensitively on the precise geometry of the structure, as will
be shown in more detail in the next section.
A measurement of this resistance change would allow one to directly
measure $\lambda$.
Calculating the relative resistance change for the wire of
Fig.~\ref{fig:system}(a) gives
\begin{equation}
\label{1Dresult}
\frac{\delta R}{R}=-1.38 \lambda,
\end{equation}
independent of ${\cal D}$ and $L$.
For silver, the estimated value of the interaction parameter is
$\lambda=+0.04$, (Ref.~\onlinecite{mota}), and therefore the resistance of
a silver wire in contact with a superconductor is reduced by 5.5\% with
respect to its normal state value.

\subsection{Generalization to arbitrary geometries} 
\label{arbitrary} 

The results obtained above are readily generalized for arbitrary
structures containing a number of normal reservoirs, two superconducting
ones and a number of diffusive wires connecting them. To illustrate
how the resistance change for such systems is calculated we consider
the geometry shown in Fig.~\ref{fig:system}(b). This is a structure
similar to the one we used recently\cite{ns} to model the experimental
setup of Ref.~\onlinecite{petrashov}. In the experiment a current was
applied along the path indicated by the arrow going from normal lead
$N_{1}$ to normal lead $N_{2}$. The voltage was measured between the
opposite normal reservoirs $N_{3}$ and $N_{4}$. The superconductors
have a phase difference $\phi$. In principle
Eqs.~(\ref{simpel}) and (\ref{diffeq}) should be solved in each branch
of the structure and the solutions matched in every nodal point.
However, it is possible to reduce this geometry to a one dimensional one
due to the symmetry of the geometry: The voltage and $f_{\rm T}$
distributions
are antisymmetric with respect to the line ${\rm S}_{1}-{\rm S}_{2}$ in
Fig.~\ref{fig:system}(b) whereas $\Delta$ and the Green function 
$\hat{g}_{\varepsilon}$ are symmetric. This allows us to consider only the
three elementary branches I, III, and IV in the calculation of
$\hat{g}_{\varepsilon}$.

For the calculation of the resistance change we need only consider the
branches I and II, each having a length $L'$, and III which has a
length $L$. As depicted in Fig.~\ref{fig:system}(c) we use
coordinates $x$, $y$ and $z$ to denote the position on branch I, II and
III, respectively. with the origins as indicated. The zeroth order
solution of Eq.~(\ref{diffeq}) is
\begin{equation}
\label{fnul}
f_{0}= \left\{
\begin{array}{ll}
L+L'-x & \mbox{(I)} \\
L & \mbox{(II)} \\
z & \mbox{(III)} 
\end{array} \right..
\end{equation}
Integrating $f_{1}''=\frac{\gamma}{{\cal D}} f_{0}$ in each branch gives
\begin{equation}
\label{feenaccent}
f'_{1}= \left\{
\begin{array}{ll}
c_{1}+\int_{0}^{x}dx'\;(L+L'-x')\;\frac{\gamma(x')}{{\cal D}}&
\mbox{(I)} \\
c_{2}+\int_{0}^{y}dy'\;L\;\frac{\gamma(y')}{{\cal D}}&\mbox{(II)} \\
c_{3}+\int_{0}^{z}dz'\;z'\;\frac{\gamma(z')}{{\cal D}}&\mbox{(III)} 
\end{array} \right..
\end{equation}
Here the integration constant $c_{1}=0$ because the current in
point $x=0$ is fixed and $c_{2}=0$ because there is no current in
branch II. Note that the situation here is different from the one
in section~\ref{simple}. There we calculated the change in current at
fixed voltage whereas in this case we calculate the voltage change at
fixed current. Consequently the resistance change is now given by
$\frac{\delta R}{R}=\frac{\delta V}{V}=\frac{f_{1}(B)}{f_{0}(B)}$.
In the nodal point A we have $f'_{1}(x=L')+f'_{1}(y=L')+
f'_{1}(z=L)=0$ and this gives
\begin{eqnarray}
\label{somisnul}
\nonumber
c_{3}&+&\int_{0}^{L} dz\;\frac{z \gamma(z)}{{\cal D}}+ 
\int_{0}^{L'} dy\; \frac{L \gamma(y)}{{\cal D}}\\
&+& \int_{0}^{L'} dx\;\frac{(L+L'-x) \gamma(x)}{{\cal D}}=0.
\end{eqnarray}
The distribution function $f_{1}$ in point A is now given by
\begin{eqnarray}
\label{feenina}
\nonumber
f_{1}(A)&=&c_{3}\int_{0}^{L} dz + \int_{0}^{L} dz \int_{0}^{z} dz'\;
\frac{z' \gamma(z')}{{\cal D}}\\
&=&c_{3} L +\int_{0}^{L} dz\;\frac{L z -z^{2}}{{\cal D}}\gamma(z).
\end{eqnarray}
Analogously we find for $f_{1}(B)$:
\begin{eqnarray}
\label{feeninb}
\nonumber
f_{1}(B)&=&f_{1}(A)-\int_{0}^{L'} dy \int_{0}^{y} dy'\;
\frac{L y' \gamma(y')}{{\cal D}}\\
&=&f_{1}(A)-\int_{0}^{L'} dy\;
\frac{y^{2} -L L'}{{\cal D}}\gamma(y).
\end{eqnarray}
Using Eq.~(\ref{somisnul}) to eliminate the first two terms in
expression~(\ref{feenina}) for $f_{1}(A)$ we calculate the resistance
change:
\begin{eqnarray}
\label{reschange}
\nonumber
\frac{\delta R}{R}&=&\int_{0}^{L'} dy\;\left(\frac{y^{2}}{L}-L-L'\right)\;
\frac{\gamma(y)}{{\cal D}}
-\int_{0}^{L} dz\; \frac{z^{2} \gamma(z)}{{\cal D}L}\\
&-&\int_{0}^{L'} dx\; \frac{(L+L'-x) \gamma(x)}{{\cal D}}.
\end{eqnarray}
The relative resistance change depends
on the phase difference $\phi$ between the superconductors through
$\gamma$, which in its turn depends on $\phi$ via the boundary conditions
imposed on $\hat{g}_{\varepsilon}$. Note that although branch IV does not
enter the calculation of the resistance change, its length does
influence the Green function, hence $\gamma$, and thus also the
resistance of the structure.

Before discussing the results of a calculation for an experimentally
relevant geometry, it is instructive to look at the (slightly unrealistic)
case of constant $\gamma$ to investigate the qualitative behavior of the
system. In this case the resistance change reduces to
\begin{equation}
\label{gammaisc}
\frac{\delta R}{R}=\frac{\gamma L^{2}}{{\cal D}} \left(
\frac{1}{3}\alpha^{3}-\frac{3}{2}\alpha^{2}-2\alpha-\frac{1}{3}
\right)~\mbox{with}~\alpha=\frac{L'}{L}
\end{equation}
which shows that, since $\gamma\sim\lambda\frac{{\cal D}}{L^{2}}$,
the relative resistance change is proportional to the interaction
parameter and independent of the absolute values of $L$, $L'$ and
${\cal D}$. Only the ratio of $L$ and $L'$ is important. In this
simple case the largest effect would be obtained for $\alpha=3.56$.
We can get an idea how sensitive the resistance change
is to the details of the geometry by comparing the cross structure
($L'=0$) with the structure in which $L'=L$.
It is easily calculated that the resistance change is higher in
the latter case by a factor of 10.5, showing that a relatively
small change in geometry causes a major modification of the resistance
change. In a realistic computation the modification due to these side
branches is not so dramatic, but is still about a factor 3.

\begin{figure}
\epsfxsize=9cm \epsfbox[18 360 577 800]{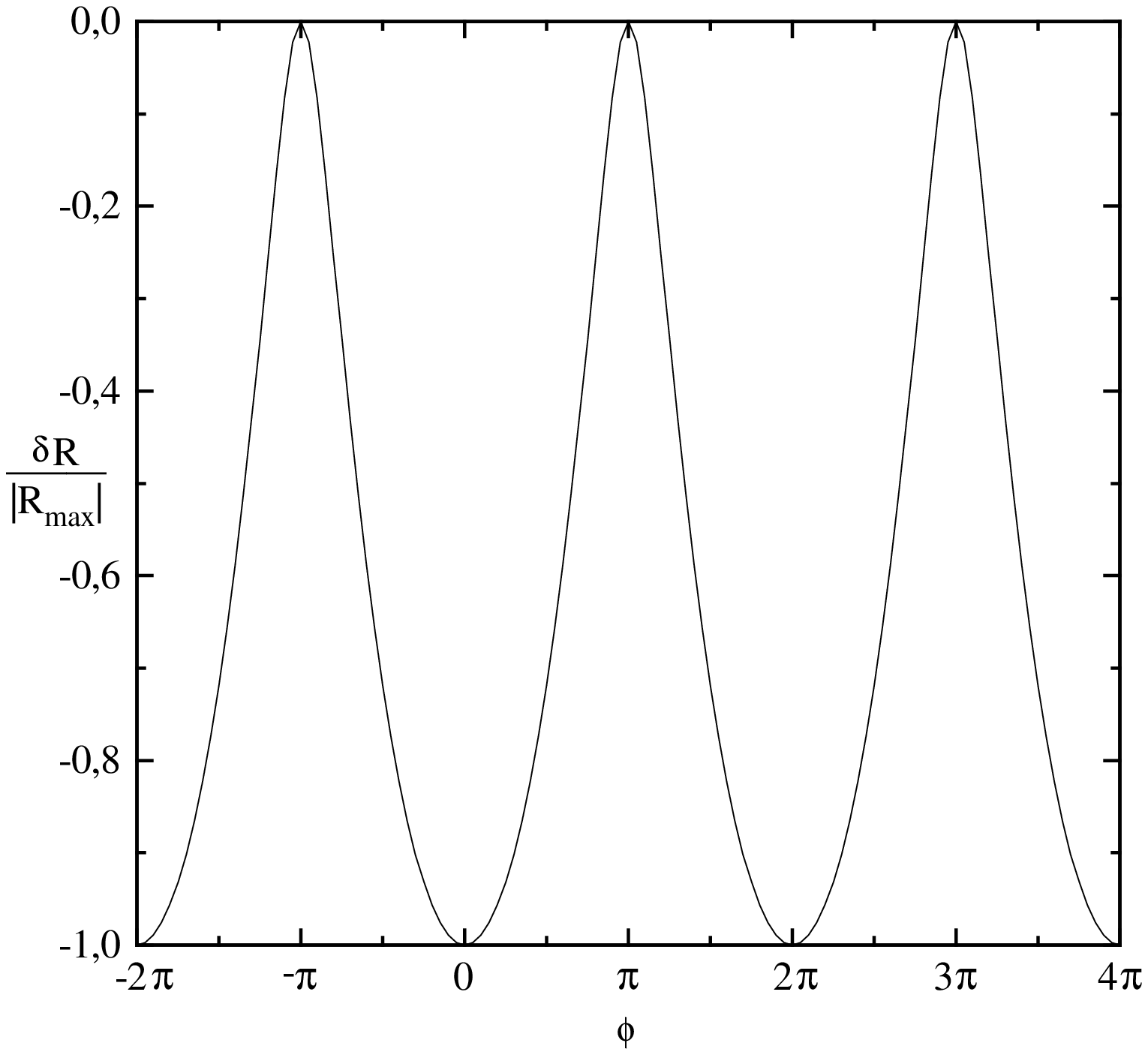}
\caption{The scaled resistance change due to the interaction effect of the
structure shown in Fig.~\protect\ref{fig:system}(b) with
$L_{I}=L_{II}=L_{IV}=L$ and $L_{III}=2L$. The magnitude of the effect
is 0.6\%.}
\label{fig:int}
\end{figure}

The result of such a realistic calculation is depicted in
Fig.~\ref{fig:int} where the scaled resistance change has been plotted
for the Andreev interferometer of Fig.~\ref{fig:system}(b) with
$L_{I}=L_{II}=L_{IV}=L$ and $L_{III}=2L$. This is the layout we used
in Ref.~\onlinecite{ns} to model the experimental setup of Petrashov et al.

There are two main differences with the previous results for 
the one dimensional wire. The first is the dependence of the 
resistance change on the phase difference $\phi$ between the 
superconductors with oscillation period $2\pi$. The effect has 
its largest (negative) value for zero phase difference and vanishes 
for a phase difference of $\pi$, when superconductivity in the  
current branch (branches I, II, and their mirror images) is 
completely suppressed.  
The second difference is the magnitude of the effect. In the one 
dimensional case this was 5.5\%. Here it is 0.6\% (for silver). 
Although this value is proportional to the interaction parameter and  
strongly depends on the length of the branches I and II, as was 
established above, a more important reason in this case is 
that the superconductors are not in the current path. As a 
consequence, the pair potential in branches I and II is smaller than 
in the one dimensional case, leading to a weaker effect. In general, 
the magnitude of the interaction effect is smaller than the influence 
of a finite temperature on the resistance. This latter phenomenon is 
the subject of the next section.

\section{The thermal effect} 
\label{thermal}

\subsection{The temperature-dependent resistance of a 1D wire}
\label{andreev}

As was shown recently by the authors,\cite{ns} the experimental 
results of Ref.~\onlinecite{petrashov} could be explained by the
temperature-dependent proximity effect. In this case
no pair potential is induced in the normal metal region by the
electron-electron interaction but coherence occurs as a result of the
finite extent $\xi = \sqrt{{\cal D}/T}$ at which the superconductivity 
penetrates into the normal metal. Although the implications
of this effect for complex structures are not always immediately
apparent, the mechanism itself is not new. It was already studied in
the middle seventies\cite{artemenko} and is in fact the
phenomenon of Andreev reflection in diffusive metals. In spite of the
fact that the effect has been known for a long time, a clear physical
picture is still lacking. 
 
To describe this effect we disregard the
second term in Eq.~(\ref{diffeq}) but we now take into account the fact
that at finite temperatures the effective diffusion coefficient is no
longer constant but depends on energy and position:
$D(\varepsilon,x)=\frac{{\cal D}}{8}
\mbox{Tr}\{{(\hat{g}^{\rm A}_{\varepsilon}(x)
+\hat{g}^{A \dagger}_{-\varepsilon}(x))^{2}} \}$. The diffusion
coefficient 
reduces to ${\cal D}$ for low and high energies. For energies 
$\varepsilon \approx {\cal D}/L^{2}$ it exhibits a maximum which is 
about twice as large as the zero energy value. The temperature 
enters the boundary conditions for Eq.~(\ref{diffeq}), but more 
importantly it determines the energy window in which the quasiparticles 
experience the energy dependence of the diffusion coefficient. 
Therefore the resistance change should vanish at both 
low and high temperatures and to have a minimum for quasiparticle 
energies $\varepsilon \approx {\cal D}/L^{2}$. 
 
\begin{figure}
\epsfxsize=8cm \epsfbox[18 310 577 750]{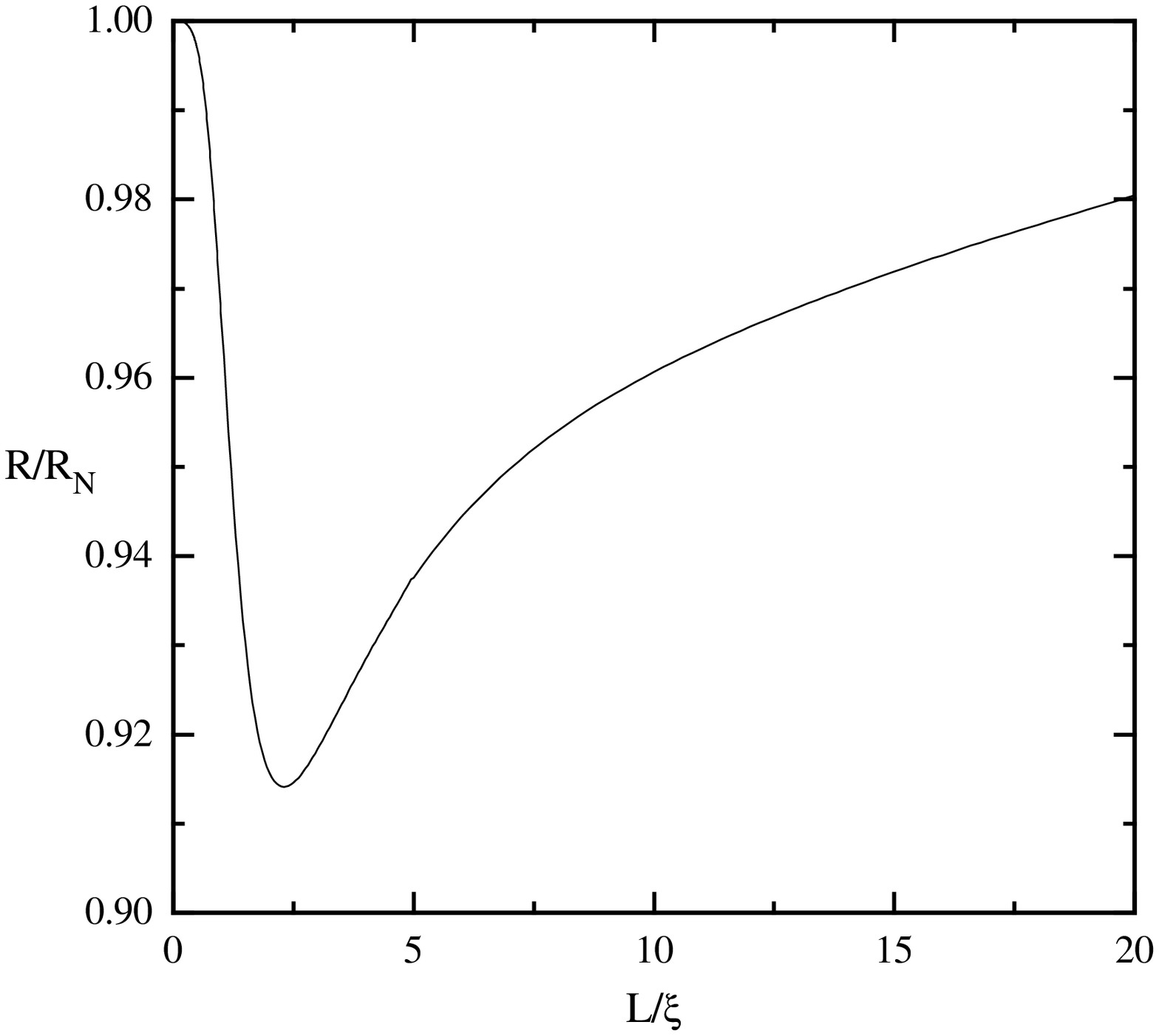}
\caption{Temperature dependence of the normalized resistance of the
structure shown in Fig.~\protect\ref{fig:system}(a). The temperature is
proportional to $(L/\xi)^{2}$.}
\label{fig:temp}
\end{figure}

As in the previous section, we will first calculate this effect for
the one dimensional structure of Fig.~\ref{fig:system}(a) and then
extend the treatment to more general structures, using the geometry
of Fig.~\ref{fig:system}(b) as an example. The normal lead is now
situated at $x=0$ and the superconducting one at $x=L$. The procedure
is as follows: We first integrate the diffusion equation for $f_{\rm T}$
twice, which gives
$f_{\rm T}(x,\varepsilon)=a(\varepsilon) \int_{0}^{x}
D^{-1}(x',\varepsilon)
dx' + b(\varepsilon)$. The integration constants may still depend on 
$\varepsilon$ and are determined by the boundary conditions 
$f_{\rm T}(0,\varepsilon)=\frac{eV}{2T}
\cosh^{-2}(\frac{\varepsilon}{2T})$
and $f_{\rm T}(L,\varepsilon)=0$. This gives
\begin{equation}
\label{disfun}
f_{\rm T}(x,\varepsilon)=\frac{eV}{2T}
\cosh^{-2}\left(\frac{\varepsilon}{2T}
\right) \left(1-\frac{m(x,\varepsilon)}{m(L,\varepsilon)}\right)
\end{equation}
with
\begin{equation}
\label{mfunc}
m(x,\varepsilon)=\frac{1}{L}\int_{0}^{x} D^{-1}(x',\varepsilon) dx'.
\end{equation}
As before, the current flowing out of the normal contact is proportional
to $f'_{\rm T}(0,\varepsilon)$ and hence we obtain for the normalized
conductance $\frac{G}{G_{N}}$:
\begin{eqnarray}
\label{normcond}
\nonumber
\frac{G}{G_{N}}&=&\frac{1}{2 {\cal D} T} \int_{0}^{\infty} d\varepsilon\;
m^{-1}(L,\varepsilon)
\cosh^{-2}\left(\frac{\varepsilon}{2T}\right)\\
&=& \frac{\xi^{2}}{2 L^{2}{\cal D}} \int_{0}^{\infty} d\tilde\varepsilon\;
m^{-1}(L,\varepsilon)
\cosh^{-2}\left(\frac{\tilde\varepsilon \xi^{2}}{2 L^{2}}\right),
\end{eqnarray}
where we have used $T={\cal D}/\xi^{2}$ and scaled the energy
with the Thouless energy $\tilde\varepsilon=\varepsilon L^{2}/{\cal D}$. 
Note that the $\cosh^{-2}(\frac{\varepsilon}{2T})$ part of the integrand 
defines the energy interval in which the quasiparticles feel the 
energy dependence of $m^{-1}(L,\varepsilon)$, as mentioned above. 
Eq.~(\ref{normcond})
shows that the normalized conductance depends on the ratio
$L/\xi \sim \sqrt{T}$ only (the factor $[{\cal D} m(L,\varepsilon)]^{-1}$
is dimensionless and does not depend on temperature).
In Fig.~\ref{fig:temp} the temperature dependence of the normalized
resistance, $\frac{R}{R_{N}}=\frac{G_{N}}{G}$, is plotted.
As expected, the effect vanishes at $T=0$ and for $T \rightarrow \infty$.
The maximum magnitude of the temperature effect is material independent
and in general larger than that of the interaction effect;
10\% vs. 5.5\% for the particular case of a silver wire.
Although the high temperature tail like the one of Fig.~\ref{fig:temp} 
was already observed in Ref.~\onlinecite{petrashov}, the low temperature 
reentrant behavior of the resistance has not yet been measured.

\subsection{Temperature effect in Andreev interferometers}
\label{gentemp}

A generalization of this theory to more complicated structures should
take into account how a resistance measurement is actually done. In the
case of the experiment of Ref.~\onlinecite{petrashov} a current was
applied between $N_{1}$ and $N_{2}$ whereupon the voltage between
$N_{3}$ and $N_{4}$ was measured (see Fig.~\ref{fig:system}).
As a consequence, there is no current
in branch II nor in its mirror image attached to reservoir $N_{4}$.
To calculate the resistance of this complex structure we proceed as
follows: We start by noting that the current is conserved for each 
quasiparticle energy $\varepsilon$. This implies that
${\cal I}_I(\varepsilon )+
{\cal I}_{II}(\varepsilon)={\cal I}_{III}(\varepsilon )$. We now
apply standard circuit theory for every quasiparticle energy (we use the
fact that due to the antisymmetry of the voltage distribution the voltage
is zero in the middle of the structure):
\begin{mathletters} 
\label{ctheory}
\begin{eqnarray}
{\cal V}_1(\varepsilon)&=&{\cal I}_{I}(\varepsilon)
{\cal R}_{I}(\varepsilon)
+{\cal I}_{III}(\varepsilon){\cal R}_{III}(\varepsilon),\\
{\cal V}_3(\varepsilon)&=&{\cal I}_{II}(\varepsilon )
{\cal R}_{II}(\varepsilon)+{\cal I}_{III}(\varepsilon)
{\cal R}_{III}(\varepsilon),
\end{eqnarray}
\end{mathletters}
where ${\cal V}_{i}(\varepsilon )=
\frac{V_{i}}{2T}\cosh^{-2}(\frac{\varepsilon}{2T})$ and the indices on
${\cal V}$ refer to the respective reservoirs and the indices on
${\cal I}$
and ${\cal R}$ refer to the appropriate branches. The proportionality
constants ${\cal R}_{I,II,III}(\varepsilon)$ in the different branches
can be deduced from the 'resistance' of the branches connecting two
reservoirs:
\begin{equation}
\label{rest}
{\cal R}_{ij}(\varepsilon)=\frac{f_{i}(\varepsilon)-
f_{j}(\varepsilon)}{f'_{i}(\varepsilon)-f'_{j}(\varepsilon)},
\end{equation} where $i$ and $j$ label the reservoirs at which the
distribution functions are evaluated.

Using the current conservation condition, Eq.~(\ref{ctheory}) can
be cast into the form of a matrix equation $\vec{{\cal V}}(\varepsilon)
={\cal R}(\varepsilon) \vec{{\cal I}}(\varepsilon)$, where
${\cal R}(\varepsilon)$ is a ($2\times2$) matrix relating the voltage and
current vectors $\vec{{\cal V}}(\varepsilon)=
({\cal V}_1(\varepsilon),{\cal V}_3(\varepsilon))^{\rm T}$ and
$\vec{{\cal I}}(\varepsilon)=
({\cal I}_I(\varepsilon),{\cal I}_{II}(\varepsilon))^{\rm T}$.
Inverting the matrix equation and integrating over all energies gives
$\vec{I}=G\vec{V}$, where $\vec{V}=(V_1,V_3)^{\rm T}$,
$\vec{I}=(I_I,I_{II})^{\rm T}$, and the conductance matrix is given by
$G=\frac{1}{2T}\int_{0}^{\infty} d\varepsilon {\cal R}^{-1}(\varepsilon)
\cosh^{-2}(\frac{\varepsilon}{2T})$. We now impose the boundary condition
mentioned above on the currents in branches I and II:
$\vec{I}=(I,0)^{T}$. This gives a relation between $V_1$ and $V_3$
in terms of the matrix elements of $G$; $V_3=-\frac{G_{21}}{G_{22}}V_1$,
which in turn must be used to calculate the experimentally measured
resistance of the structure:
$R=\frac{2V_{1}}{I}=2(G_{11}-\frac{G_{12}G_{21}}{G_{22}})^{-1}$.
The resistance depends again on the phase difference between the
two superconducting reservoirs because of the boundary conditions 
imposed on the Green function.

\begin{figure}
\epsfxsize=9cm \epsfbox[18 360 577 780]{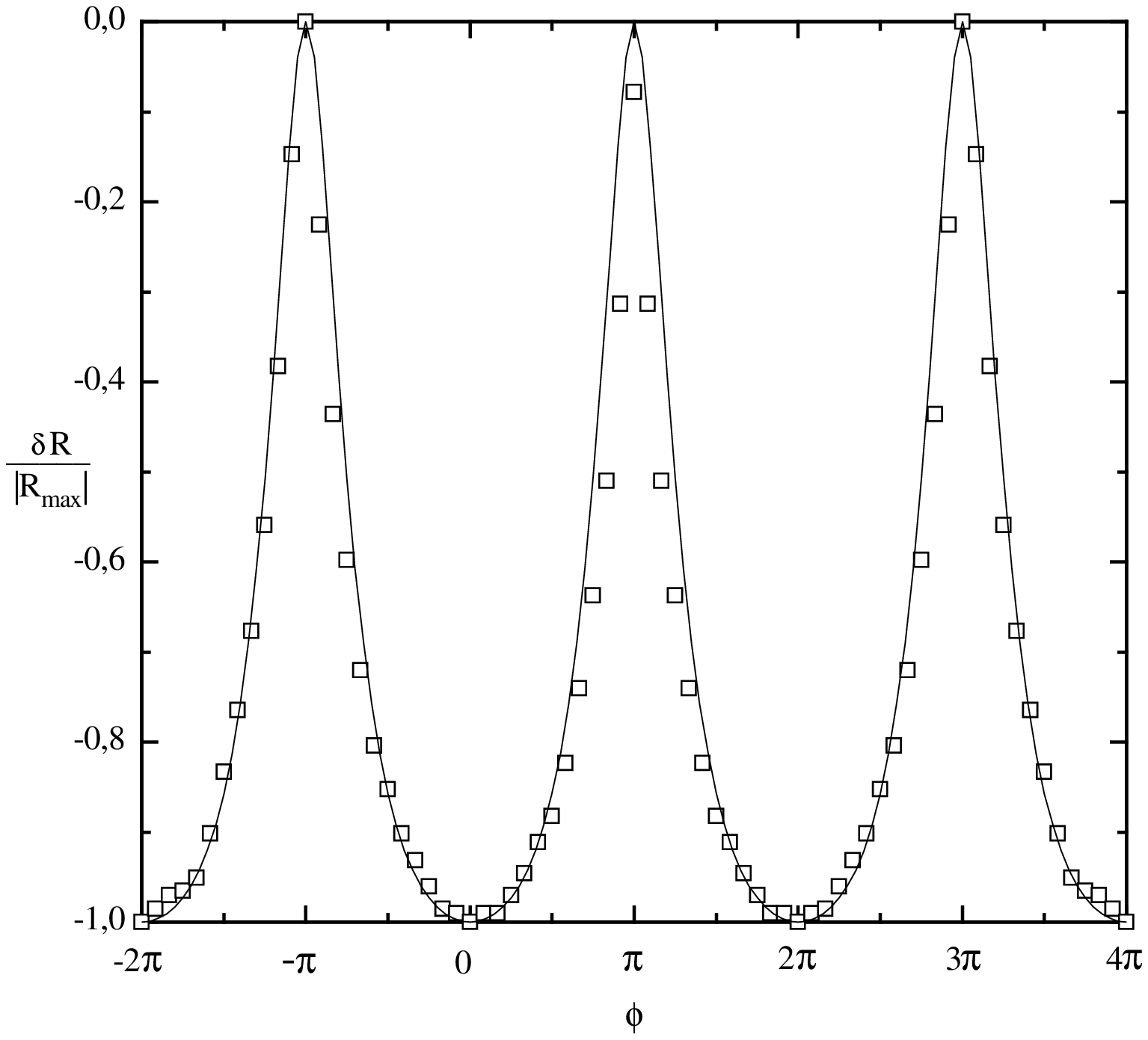}
\caption{Scaled resistance change due to the thermal effect of the
structure shown in Fig.~\protect\ref{fig:system}(b) with
$L_{I}=L_{II}=L_{IV}=L$ and $L_{III}=2L$ for $L/\xi=3$. The magnitude
of the effect is 9.7\%. The squares are the experimental data of
Ref.~\protect\onlinecite{petrashov}.}
\label{fig:temp2}
\end{figure}

In Fig.~\ref{fig:temp2} we have plotted the results of this calculation
for the previously used geometry of Fig.~\ref{fig:system}(b) with
$L_{I}=L_{II}=L_{IV}=L$ and $L_{III}=2L$. This particular calculation
was done for the value of $L/\xi=3$ which gave the biggest effect.
Also shown are the experimental results of Ref.~\onlinecite{petrashov}.

Again we find $2\pi$ periodic oscillations like in Fig.~\ref{fig:int}, 
but although the phase dependence looks similar there are some distinct 
differences. First of all the shape of the oscillations is different, 
especially near the minima, where the thermal effect produces less 
narrow peaks than the interaction effect. The amplitude of the thermal 
oscillations is material independent whereas that of Fig.~\ref{fig:int} 
is proportional to the interaction parameter of the metal. Moreover, 
the maximal magnitude of the oscillations, 9.7\% in this particular case, 
are in general much larger in the thermal effect, which makes observation 
of the previously discussed oscillations more difficult. 
 
We obtain an excellent fit with the experimental data in 
Fig.~\ref{fig:temp2}, where the magnitude of the oscillations is about
11\%. However, a decisive check on our theory would be 
provided by the observation of the temperature dependence of the 
resistance like the one plotted Fig.~\ref{fig:temp} for the system 
considered here. Indeed, unpublished data by the authors of 
Ref.~\onlinecite{petrashov} also show a remarkable 
agreement\cite{delsing} with our calculations.\cite{ns} However, 
the reentrant behavior has not yet been observed unambiguously. 

\section{Summary} 
\label{end}

We have performed calculations of the distribution of the electrostatic
and nonequilibrium chemical potential in a one dimensional disordered 
superconducting hybrid wire. We showed that the two behave differently as
a function of temperature and that they are in fact only equal in the high
temperature limit. We have proposed an experimental setup to measure these
different potential distributions.
We have also presented a computation for this wire using a recently
developed\cite{ns} mechanism, that causes the resistance of a diffusive
superconducting hybrid structure to change at zero temperature. The latter
is in contrast with the well known thermal mechanism of Andreev reflection
in diffusive metals, where the resistance change vanishes for low
temperatures.
In addition to this new result we have given a detailed account of
the calculation performed in Ref.~\onlinecite{ns} for the experimental
setup of Ref.~\onlinecite{petrashov}.
Since the relative resistance change due to the novel mechanism is
proportional to the interaction parameter in the normal metal, observation
of this effect would allow a direct measurement of this physical quantity.
Furthermore, we have shown how to calculate the resistance of an arbitrary
Andreev interferometer using finite-temperature proximity effect theory.
Because the finite temperature effect generally causes a much larger
resistance change, this is the correct theory to describe an experiment
like the one performed in Ref.~\onlinecite{petrashov}.

\acknowledgments
The authors would like to thank Daniel Esteve, Gerrit Bauer, Luuk Mur,
Mark Visscher, and Henk Stoof for valuable discussions.
This work is part of the research program of 
the "Stichting voor Fundamenteel Onderzoek der Materie" (FOM), which is 
financially supported by the "Nederlandse Organisatie voor
Wetenschappelijk Onderzoek" (NWO).

\end{document}